\documentclass[12pt]{article}
\usepackage{graphicx}
\usepackage{amssymb,amsmath,amsfonts,palatino,amsthm}
\usepackage{amssymb}
\usepackage{epstopdf}
\newcommand{\beq}{\begin{equation}}
\newcommand{\eeq}{\end{equation}}

\newcommand{\f}{\begin{equation}}
\newcommand{\ff}{\end{equation}}

\begin{document}

%%%%%%%%%%%%%%%%%%%%%%%%%%%%%%%%%%%%%%%%%%%%%%%%
\title{The quantization of unimodular gravity and the cosmological constant problems}
\author{Lee Smolin\thanks{lsmolin@perimeterinstitute.ca}
\\
\\
Perimeter Institute for Theoretical Physics,\\
31 Caroline Street North, Waterloo, Ontario N2J 2Y5, Canada}
\date{\today}
\maketitle

\begin{abstract}
A quantization of unimodular gravity is described, which results in a quantum effective action which is also unimodular, ie a function of a metric with fixed determinant.  A consequence is that contributions to the energy momentum tensor 
of the form of $g_{ab} C$, where $C$ is a spacetime constant, whether classical or quantum, are not sources of curvature
in the equations of motion derived from the quantum effective action.   This solves the first cosmological constant problem,
which is suppressing the enormous contributions to the cosmological constant coming from quantum corrections.  

We discuss several forms of uniodular gravity and put two of them, including one proposed by Henneaux and Teitelboim, in constrained Hamiltonian form.  The path integral is constructed from the latter.

Furthermore, the second cosmological constant problem, which is why the measured value is so small, is also addressed
by this theory.  We argue that a mechanism first  proposed by Ng and van Dam for suppressing the cosmological constant
by quantum effects  obtains  at the semiclassical level.

\end{abstract}
\newpage

\tableofcontents

\section{Introduction}

The problem of the cosmological constant constitutes at least three puzzles,  1) Why is $\Lambda$ not of the order of $M_{Pl}^2$ due to quantum corrections?  2)  Why after cancellation of this term as well as possible large contributions from symmetry breaking, is the measured value of the dark energy  $10^{-120} M_{Pl}^2$?  3) Is it a coincidence that this is presently roughly a factor of $3$ times
the average matter density?  

One approach to the first problem, which must have come to the mind of many researchers, is that in nature the gravitational field
just does not couple to vacuum energy.  However, an objection to this quickly raises its head, which is that we can argue from the
very precise tests which have been done on the equivalence principle that energy from quantum corrections does fall the same way
as other energy. 

But suppose what one means by this is not precisely that gravity does not couple to quantum corrections to energies but something
else, which is that terms in the energy momentum tensor proportional to the metric $g_{ab}$ do not come into some modified
form of the Einstein equations.  That is, suppose the equations of motion of the metric are unchanged under a modification
of the energy momentum tensor of the form,
\f
T_{ab} \rightarrow T^\prime_{ab} = T_{ab}+ g_{ab} C
\label{modify}
\ff
where $C$ is a spacetime constant. Then, in particular, vacuum corrections to the cosmological constant, which are of this
form, would not affect the curvature of spacetime.  This could be true in a theory where nonetheless vacuum fluctuations  gravitate when they contribute to the mass of a bound state, which having a specific rest frame is never of the form of $g_{ab} C$.  

Afshordi\cite{niayesh}  has recently discussed an approach to solving the first cosmological constant problem which is of this form.  The logic of his approach is very simple.   Afshordi
proposes that the fact that such a term in the energy momentum tensor proportional to $g_{ab} C$ is not observed means that general relativity must be modified so the source term on the right hand side 
of the gravitational field equations is  $T_{ab} -\frac{1}{4} g_{ab} T$ . 
 
One might expect that this would lead immediately to inconsistencies with the
successful predictions of general relativity, but he  finds a very clever way to implement this so that the basic predictions of 
general relativity are not affected.    Other approaches of this kind, that {\it degravitate} the zero point energy at appropriate scales, have been discussed by \cite{degrav}.  

Here we note that a theory making use of this mechanism is almost as old as general relativity.   This is unimodular gravity, first written
down by Einstein\cite{einsteinuni} in 1919.  
Unimodular gravity\cite{einsteinuni}-\cite{bombelliuni}  modifies general relativity by imposing a constraint that the metric of spacetime, ${g}_{ab}$,  have a fixed determinant.  This has the effect of reducing the gauge symmetry from full spacetime diffeomorphism invariance to invariance only under diffeomorphisms that preserve this non-dynamical fixed volume element. In spite of these differences,  the field equations are the same as general relativity.  Only now the cosmological constant, $\Lambda$, is a constant of integration rather than a parameter of the lagrangian.  

If one asks why this kind of approach to the cosmological constant problem has not been more fully considered, part of the reason is that the quantization of unimodular gravity has remained obscure.  For example, as discussed by Unruh in \cite{unruhuni}, there are additional constraints that complicate the construction of the quantum theory.  But  as the first cosmological constant problem concerns suppressing large quantum corrections, it must be solved in the context of a quantum theory.  This means that the symmetry (\ref{modify}) has to be satisfied by the full quantum equations of motion, which follow from the quantum effective action.  However, as Weinberg pointed out, it is not clear whether there is any theory whose quantization yields a quantum effective action which is a functional of the 
unimodular metric\cite{weinberguni}.   This question is
resolved affirmatively here.  

To give a well defined construction of the path integral for unimodular gravity, we  follow Henneaux and Teitelboim\cite{HTuni}  in making a background density, which is not always
written down in discussions of unimodular gravity, but must be there for the action principle to be  sensible,
 into a dynamical field.  

In the next section we review two approaches to unimodular gravity which are in the literature, based on different action 
principles.    In the first of these the restriction to metrics with fixed determinants is imposed on the action, while in the 
second, due to 
Henneaux and Teitelboim\cite{HTuni}, it appears as an equations of motion.  (Two more forms of unimodular gravity are described in an appendix.)  This is useful because we can take advantage of these different forms, which 
agree at the classical level, to find one most amenable to quantization.  

We put the two forms of the  theory in constrained hamiltonian form in section 3.  The quantization of the Henneaux and Teitelboim form, \cite{HTuni}, is the subject of section 4.  We construct the gauge fixed, configuration space partition function, following the usual procedure for gauge theories, starting with the constrained Hamiltonian form of the path integral.   The goal of that section is a single result, which is that the condition that the metric have fixed determinant is maintained under quantization of the theory.  This means that even after quantum corrections are included the quantum effective action is a functional of a metric with fixed determinant.  This implies that the
equations of motion that follow from the quantum effective action have the symmetry (\ref{modify}).  As this means that even in the quantum theory contributions to the energy momentum tensor of the form $g_{ab} C$ do not couple to the metric, this appears to solve the first cosmological constant problem.    

In section 5 we discuss the second cosmological constant problem and show that a mechanism originally proposed by
Ng and van Dam \cite{nguni} is realized at the level of the semiclassical approximation.  And we find a bonus, which is an indication that the theory addresses the third, coincidence problem as well.  However, these considerations are still more formal and also depend
on assumptions about the measurement theory of quantum cosmology, a famously wooly subject. 

We close with some remarks on the possible implications of these results.

\section{Variations on the theme of  unimodular gravity}

We now describe two different realizations of the idea of unimodular gravity\footnote{Two more are described in the appendix.}.

\subsection{The original idea}

Unimodular gravity is sometimes written as the reduction of the Einstein action\footnote{We assume throughout that the
spacetime is spatially compact.}
\f
S^{Einstein} = \int_{\cal M} \sqrt{-g} \left  (  -\frac{1}{8\pi G} g^{ab} R_{ab} + {\cal L}^{matter} \right )
\label{Einstein}
\ff
to the case that the metric tensor $g_{ab}$ is restricted by the unimodular condition
\f
g= det( {g}_{ab}) =1
\label{uni1}
\ff
This of course is imprecise as (\ref{uni1}) equates a density with a function.  What is meant is, more precisely,
that there is a fixed density $\epsilon_0$  so that 
\f
\sqrt{-\bar{g}} =\epsilon_0
\label{uni2}
\ff
The action is then 
\f
S^{uni} = \int_{\cal M} \epsilon_0 \left (  -\frac{1}{8\pi G}  \bar{g}^{ab} R_{ab} + {\cal L}^{matter} (\bar{g}_{ab}, \psi ) \right )
\label{uniaction1}
\ff
Here $\psi$ refers to matter fields, and by $\bar{g}_{ab}$ we mean the metric restricted by {\ref{uni2}).  
This implies a breaking of spacetime diffeomophism invariance as the invariance group of (\ref{uniaction1}) is
the volume preserving diffeomoprhisms that preserve $\epsilon_0$.  These are generated by vector fields $v^a$\
that satisfy
\f
\partial_a (\epsilon_0 v^a) =0
\label{uni3}
\ff
The equations of motion are tracefree because the variation of $g_{ab}$ is done subject to the constraint
(\ref{uni2}).  
\f
 R_{ab}-\frac{1}{4}\bar{g}_{ab} R = 
4 \pi G \left ( E_{ab}-\frac{1}{4}\bar{g}_{ab} E \right )
\label{EOM1}
\ff
The matter sources are represented by
\f
E_{ab} \equiv \frac{\delta {\cal L}^{matter}}{\delta g_{ab}}
\ff
Note that this differs from the usual covariantly conserved
\f
T_{ab} \equiv \frac{1}{\sqrt{-g}} \frac{\delta \sqrt{-g} {\cal L}^{matter}}{\delta g_{ab}}
\ff
In fact we have 
\f
T_{ab}= E_{ab}-\frac{1}{2}g_{ab} {\cal L}^{matter}
\ff
so that the traces satisfy 
\f
T= E-2 {\cal L}^{matter}
\ff
Note that
\f
T_{ab}-\frac{1}{4}g_{ab}T = E_{ab}-\frac{1}{4}g_{ab}E
\ff
The divergence of the equations of motion yeilds
\f
\partial_a \left ( R+4\pi G T      \right )=0
\ff
which allows us to define a constant of integration, $\Lambda$, by
\f
R +4 \pi G T= -4\Lambda
\label{Lambda1}
\ff
after which we can rewrite the equations of motion as the Einstein equations for an arbitrary constant $\Lambda$
\f
G_{ab}-\Lambda g_{ab} =4\pi G T_{ab}
\ff
As promised, the spacetime curvature doesn't couple to corrections to the energy-momentum tensor of the 
form\footnote{But as we see above, the addition of terms of the form of functions times $g_{ab}$ does matter.}
of $g_{ab} C$.  
To see this note that the equations of motion (\ref{EOM1}) and (\ref{EOM2}) are unchanged under
any modification of the form (\ref{modify}).  Indeed, this implies from (\ref{Lambda1}) that at the same time we must
shift
\f
\Lambda \rightarrow \Lambda -4 \pi G C
\label{modify2}
\ff
so that under the combination (\ref{modify}) and (\ref{modify2}) the Einstein equations, (\ref{EOM3}) are unchanged.

\subsection{The Henneaux-Teitelboim formulation of unimodular gravity}

Henneaux and Teitelboim\cite{HTuni} reformulated unimodular gravity so that the action depends on the full unconstrained 
metric and the gauge symmetry includes the full diffeomorphism group of the manifold. 
They do this as follows.  They introduce two auxiliary fields.  The first 
is a three form  
$a_{abc}$, whose field strength is $b_{abcd}=da_{abcd}$.  The dual is a density\footnote{We use the notation where tildes
refer to densities.}
\f
\tilde{b}= \frac{1}{4!} \epsilon^{abcd} b_{abcd}= \partial_a \tilde{a}^a
\ff
where $\tilde{a}^a$ is the vector density field defined as $\tilde{a}^a= \frac{1}{6} \epsilon^{abcd}a_{bcd}$.
The second is scalar field $\phi$ which serves as a lagrange multiplier.  They then write the action
\f
S^{HT} (\bar{g}_{ab}, a_{abc},\phi,...)  = \int_{\cal M} \sqrt{-g} \left ( -\frac{1}{8\pi G}  (  \bar{g}^{ab} R_{ab} + \phi)  + {\cal L}^{matter} \right )
+ \frac{1}{8\pi G}  \phi \tilde{b}
\label{HT}
\ff
By varying $\phi$ they find the unimodular condition emerging as an equation of motion
\f
\sqrt{-g}= \tilde{b}
\label{uni5}
\ff
Varying $\tilde{a}^a$ they find
\f
\partial_a \phi =0
\label{HT1}
\ff
so that the field $\phi$ becomes a spacetime constant on solutions, so we can write 
\f
\phi (x) = \Lambda
\ff
Varying $g^{ab}$ we find the Einstein equations for any value of the constant $\Lambda$
\f
G_{ab}-\Lambda g_{ab} = 4 \pi G T_{ab}
\label{HT2}
\ff
We note that (\ref{uni5}) implies that the metric covariant derivative satisfies
\f
\nabla_a \tilde{b}= \nabla_a \sqrt{-g}=0
\label{critical}
\ff
Note that the decoupling of the terms in the energy momentum tensor of the form of  $g_{ab} C $ still occurs, but it is said differently.
The equations of motion (\ref{uni5},\ref{HT1},\ref{HT2}) are invariant under the simultaneous shift
\f
T_{ab}\rightarrow T_{ab} + g_{ab} C, \ \ \ \  \phi \rightarrow \phi - 4 \pi G C
\label{HTshift1}
\ff
Finally,  we can give an interpretation of the field $\tilde{a}$ as follows.  Let us integrate 
(\ref{uni5}) over a region of spacetime $\cal R$ bounded by two spacelike surfaces $\Sigma_1$ and $\Sigma_2$.  Then we have
\f
\int_{\Sigma_2} a - \int_{\Sigma_1} a = Vol = \int_{\cal R} \sqrt{-g}
\ff
That is $a$ pulled back into the surface is a time coordinate that measures the total four volume to the past of that surface.  We can consider that time coordinate associated to a surface $\Sigma$ to be $T= \int_{\Sigma} a $.

\section{Constrained hamiltonian dynamics of unimodular gravity}

We now study the hamiltonian dynamics and quantization of the different forms of unimodular gravity.

\subsection{Hamiltonian dynamics of the original unimodular gravity}

We start with the pure unimodular theory given by (\ref{uni2}).  

It is straightforward to perform the Hamiltonian analysis.  The canonical momenta are,
\f
\tilde{\pi}^{ij}= \epsilon_0 ( K^{ij}-q^{ij} K    )
\label{mom1}
\ff
We use the unimodular condition (\ref{uni2}) to express
\f
N=\frac{\epsilon_0}{\sqrt{q}}
\ff

The hamiltonian is
\f
H= \int_\Sigma \left ( \epsilon_0  (h_0+ 4 \pi G \rho )  + N^i {\cal D}_i    \right )
\ff
where $h_0$ is related to the usual Hamiltonian constrant.
\f
h_o = \frac{1}{\epsilon_0^2} \left [\tilde{\pi}^{jk}  \tilde{\pi}_{jk} -\frac{1}{2} \tilde{\pi}^2   \right ] + {}^{3}R 
\ff
where $\rho$ contains terms in matter fields and   ${\cal D}_i$ are the usual generator of diffeomorphisms, which 
are first class constraints
\f
{\cal D}_i= \tilde{\pi}^{jk}{\cal L}_i q_{jk} =0
\label{diffeos}
\ff
There is at first
no Hamiltonian constraint.  But there are new constraints that come from the preservation of of the ${\cal D}_i$ by the Hamiltonian, which are
\f
{\cal S}_i =\partial_i (h_o + 4 \pi G \rho )
\label{Si}
\ff

The ${\cal S}_i$ are locally one constraint, because
\f
\nabla_{[j}{\cal S}_{k]} =0 . 
\ff
This means that if we smear the constraint with a vector density, $\tilde{w}^i$ so
\f
{\cal S} (\tilde{w}^i ) = \int_\Sigma \tilde{w}^i {\cal S} _i
\ff
Then we have an equivalence relation
\f
\tilde{w}^i  \rightarrow \tilde{w}^{i \prime} = \tilde{w}^i + \partial_j \tilde{\rho}^{ij} 
\ff
where $ \tilde{\rho}^{ij} = -  \tilde{\rho}^{ji} $.  

We add the constraints $S_i$ to the Hamiltonian with lagrange multipliers $\tilde{w}^i$ to find 
\f
H= \int_\Sigma \left ( (\epsilon_0 + \partial_i \tilde{w}^i ) (h_o + 4 \pi G \rho )
 + N^i {\cal D}_i    \right )
\ff
 It is easy to check that ${\cal D}_i$ and  ${\cal S}_i$ form a first class
algebra of constraints.  

\subsection{Hamiltonian dynamics of the Henneaux-Teitelboim form}

As an alternative we consider the Hamiltonian constraint analysis of the Henneaux-Teiltelboim form of the theory (\ref{HT}).  We do the usual $3+1$ decomposition and define momenta for all the fields.  We
find, besides the usual
\f
\tilde{\pi}^{ij}= \sqrt{q}( k^{ij}-q^{ij} K)
\ff
also the primary constraints 
\f
P_i =P_\phi =\pi_N = \pi_{N^i} =0
\ff
and 
\f
{\cal E}= P_0 + \phi =0
\ff
Here $P_a$ are the momenta conjugate to the $\tilde{a}^a$.  
We form the Hamiltonian and, taking into account the preservation of $\pi_{N^i} $), find the usual spatial
diffeomorphism constraints (\ref{diffeos}).  The hamiltonian at this stage is,
\f
H= \int_\Sigma \left ( \tilde{\pi}^{jk}\dot{q}_{jk} + P_a \dot{\tilde{a}}^a - N\sqrt{q}(\frac{1}{8\pi G} [R+ \phi ] + \rho ) 
+\phi \partial_a \tilde{a}^a + \alpha {\cal E} + N^i {\cal D}_i
\right )
\label{ham1}
\ff
where we have introduced a lagrange multiplier $\alpha$ to express the constraint ${\cal E}$.  
As usual the preservation of $\pi_N$ results in the Hamiltonian constraint
\f
{\cal H}= h_0 +\phi +4 \pi G \rho =0
\ff
where $h_0$ is the usual Hamiltonian constraint in the form of a density of weight zero, but now expressed as
\f
h_o = \frac{1}{q} \left [\tilde{\pi}^{jk}  \tilde{\pi}_{jk} -\frac{1}{2} \tilde{\pi}^2   \right ] + {}^{3}R 
\ff
We now
compute $\{ H, P_\phi \}$ which generates a secondary constraint
\f
{\cal C}= N\sqrt{q}- \partial_i \tilde{a}^i
\ff
This together with $\pi_N=0$ are second class
\f
\{ {\cal C}, \pi_N \}=\sqrt{q} .
\ff
We then eliminate ${\cal C}$ by solving it for $N$ and replacing everywhere
\f
N=\frac{\partial_i \tilde{a}^i}{\sqrt{q}}
\ff
and at the same time eliminating $\pi_N$.  We also eliminate the pair $\phi$ and $P_\phi$ by everywhere 
using  ${\cal E}$.  Using ${\cal E}$ in ${\cal H}$ results in the constraint
\f
{\cal W}= P_0 -h_0 -4 \pi G \rho =0
\label{W}
\ff

The result is a theory with variables in canonical pairs, $(q_{jk}, \tilde{\pi}^{ij})$ and $(\tilde{a}^a, P_a)$
with a Hamiltonian
\f
H = \int_\Sigma \left ( -(\partial_i \tilde{a}^i)(h_0 +4 \pi G \rho ) +N^i {\cal D}_i 
\right )
\label{ham2}
\ff
and constraints $P_i$ and ${\cal W}$.  The latter is preserved by the evolution generated by $H$ while 
preservation of $P_i$ leads to a new set of constraints
\f
{\cal S}_i = \partial_i (h_0+4 \pi G \rho ) =0
\ff
It is easy to check that these form a first class algebra with the ${\cal D}_i$.  We can express these with 
a lagrange multiplier $\tilde{w}^i$ and the constraints $P_i$ with lagrange multipliers $\tilde{u}^i $ so we have finally
\f
H = \int_\Sigma \left ( -(\partial_i \tilde{a}^i)(h_0 +4 \pi G \rho ) +N^i {\cal D}_i +\tilde{w}^i {\cal S}_i + \tilde{u}^i P_i 
\right )
\label{ham3}
\ff
Note that the constraint ${\cal H}$ has been eliminated in favor of the ${\cal W}$ and the ${\cal S}_i$.  There is in fact only one of the
latter per point because they satisfy 
\f
\partial_{[j} {\cal S}_{k]} =0
\ff
Note that the constraints $P_i$ generate
\f
\delta \tilde{a}^i = \{ \tilde{a}^i , \int_\Sigma \tilde{u}^i P_i \} =  \tilde{u}^i 
\ff
so the $\tilde{a}^i $ are pure gauge.  On the other hand,  $\tilde{a}^0$ defines a natural time coordinate.  Define
\f
{\cal P} = \int_\sigma \sqrt{q} P_0 
\ff
Then we have from $\cal W$, (\ref{W}),
\f
{\cal P} = H_0= \int_\sigma \sqrt{q} ( h_0 +4 \pi G \rho )
\label{ham}
\ff
So we see that $\cal W$ generates changes in $\tilde{a}^0$ of the form of 
\f
\delta \tilde{a}^0 = \tau \sqrt{q}
\ff
and that 
these are correlated with evolution in the other variables generated by the Hamiltonian $H_0$.  Since there is no more a Hamiltonian constraint, the equation (\ref{ham}) can be seen as giving evolution in time, as measured by changes in total four volume, as discussed above.

\section{Construction of the path integral quantization}

The two versions of unimodular gravity we presented in section 2 have the same equations of motion, which 
are Einstein's equations with an arbitrary cosmological constant.  
As the unimodular condition is imposed in the first action and found as an equation of motion in the second, the quantum
theories may be expected to be different.  We will discuss here only the quantization of the latter,  Henneaux-Teitelboim  formulation, which we will find has the property that the quantum effective action is also an action for unimodular gravity.

\subsection{The path integral for the Henneaux-Teitelboim formulation}

We follow the standard construct the path integral from the the Hamiltonian quantum dynamics.  The partition 
function is
\begin{eqnarray}
Z &=&  \int  dq_{ij} d \tilde{\pi}^{kl} d\tilde{a}^a dP_a   d\Psi  dP_\Psi   \delta ({\cal D}_i ) \delta ({\cal S}_i ) \delta ({\cal W}) \delta (P_i )
\delta (\mbox{ gauge fixing})  Det_{FP}
\nonumber \\
 & &\times  exp \  \imath \int dt \int_\Sigma \left (   \tilde{\pi}^{jk}\dot{q}_{jk}  + P_a \dot{\tilde{a}}^a +(\partial_i \tilde{a}^i)(h_0 +4 \pi G \rho )    \right ) 
 \label{partition1}
\end{eqnarray}
where $\Psi$ are generic matter degrees of freedom and $P_\Psi$ are their conjugate momenta. 

Regarding the gauge fixing conditions, there are are eight degrees of freedom in the gauge invariance to constrain.  Four are the usual spacetime diffeomorphisms, which are unconstrained in the Henneaux-Teitelboim form of the theory. These are generated by
the ${\cal D}_i$ and the (one mode per point) present in the ${\cal S}_i$.  The other four are the gauge invariance
$\tilde{a}^a \rightarrow \tilde{a}^a + \tilde{w}^a$ generated by the first class constraints $P_i$ and the ${\cal E}$.  We could fix the gauge immediately, but for generality we will for a few steps leave the gauge fixing unspecified, except to assume that the gauge fixing functions do not involve the momenta.  

We next perform the integration over $P_0$ which uses up the delta function in $\cal W$, and the integration 
over $P_i$, which is trivial.  At the same time we 
 express the delta functions for the first class constraints ${\cal D}_i$ and  ${\cal S}_i$ by integration over lagrange
multipliers $N^i$ and $\tilde{w}^i$.  As the manifold is assumed spatially compact,  we neglect boundary terms to find
\begin{eqnarray}
Z &=&  \int  dq_{ij} d \tilde{\pi}^{kl} d\tilde{a}^a dN^i d\tilde{w}^i d\Psi dP_\Psi
\delta (\mbox{ gauge fixing})  Det_{FP}
\nonumber \\
 & &\times  exp \  \imath \int dt \int_\Sigma \left (   \tilde{\pi}^{jk}\dot{q}_{jk}   +(\partial_a \tilde{a}^a+ \partial_i \tilde{w}^i)(h_0 +4 \pi G \rho )  -N^i {\cal D}_i \right ) 
 \label{partition2}
\end{eqnarray}

We next shift
\f
\tilde{a}^i \rightarrow \tilde{a}^i - \tilde{w}^i 
\ff
so that the integrand no longer depends on $\tilde{w}^i $, which we then integrate trivially.  We find 
\begin{eqnarray}
Z &=&  \int  dq_{ij} d \tilde{\pi}^{kl} d\tilde{a}^a dN^i d\Psi dP_\Psi
\delta (\mbox{ gauge fixing})  Det_{FP}
\nonumber \\
 & &\times  exp \  \imath \int dt \int_\Sigma \left (   \tilde{\pi}^{jk}\dot{q}_{jk}   +(\partial_a \tilde{a}^a )(h_0 +4 \pi G \rho )  -N^i {\cal D}_i \right ) 
 \label{partition3}
\end{eqnarray}
We can now perform the integration over the momenta $\tilde{\pi}^{jk}$ and $P-\Psi$.  The steps that follow are the same as in the usual case,
except that we have in (\ref{partition3}), in place of $N$ the ratio of  $\tilde{b}/\sqrt{q}$.    Recalling that $N$ is in the definition of $K$ as in 
$K_{ij}= \frac{1}{N} (\dot{q}_{ij} +...) $ we find
\begin{eqnarray}
Z &=&  \int  dq_{ij}  dN^i d\tilde{a}^a d\Psi
\delta (\mbox{ gauge fixing})  Det_{FP}
\nonumber \\
 & &\times  exp \  \imath \int dt \int_\Sigma \left ( \tilde{b} ( \frac{1}{8\pi G}\bar{R} +{\cal L}^{matter} ) \right ) 
 \label{partition4}
\end{eqnarray}
where $\bar{R}$ is the Ricci scalar computer with the constraint that part of the metric is fixed in terms of $\tilde{b}$ by 
\f
N=\frac{\tilde{b}}{\sqrt{q}}
\ff

Indeed, the path integral (\ref{partition4}) is a function only of nine out of ten components of the metric, the integral over the lapse is missing.  We can put it back by introducing unity in the form
\f
1 =\int dN \delta (N- \frac{\tilde{b}}{\sqrt{q}}   )
\ff
which up to a factor in the measure gives us
\begin{eqnarray}
Z &=&  \int  dg_{ab} d\tilde{a}^a d\Psi \delta (\sqrt{-g} -\tilde{b} )
\delta (\mbox{ gauge fixing})  Det_{FP} \sqrt{-g} 
\nonumber \\
 & &\times  exp \  \imath \int dt \int_\Sigma \left ( \tilde{b} ( \frac{1}{8\pi G}R +{\cal L}^{matter} ) \right ) 
 \label{partition5}
\end{eqnarray}
Now we see that the path integral is an integration over metrics constrained by the unimodular condition in the form (\ref{uni5}).

\subsection{Gauge fixing}

We now can introduce gauge fixing conditions.  Among the simple gauge fixing conditions we might try are 
\f
\tilde{f}_i = \tilde{a}_i =0  , \ \ \ \ \tilde{f}_0 = \tilde{a}_0 - t \epsilon_0 =0
\ff
for some function $t$, which will become the effective time coordinate.  $\epsilon_0$ is a fixed density needed to make the 
density weights balence out. Then we have  
\f
\tilde{b}=\epsilon_0.  
\ff
The path integral now becomes
\begin{eqnarray}
Z &=&  \int  dg_{ab} d\Psi  \delta (\sqrt{-g} -\epsilon_0 )
\delta (\mbox{ gauge fixing})  Det_{FP} \sqrt{-g} 
\nonumber \\
 & &\times  exp \  \imath \int dt \int_\Sigma \left ( \epsilon_0 ( \frac{1}{8\pi G}R +{\cal L}^{matter} ) \right ) 
 \label{partition5b}
\end{eqnarray}

so we return to the action of the original form of unimodular gravity, $S^{uni}$, from  (\ref{uniaction1}).

\subsection{Derivation of the quantum effective action}

We can now define a perturbative expansion of the partition function of the form (\ref{partition5b}).  To do this we re-express the theory as an expansion around 
a background spacetime 
\f
g_{ab}= g_{a}^{0  \ c } X_{cb}
\ff
with $det(g^0) =\epsilon_0$.  We then write
\f
\bar{X}_{ab} =  [exp(h_{..})]_{ab}
\ff
where the perturbative field,  $h_{ab}$,  is tracefree,
\f
\delta^{ab}h_{ab} =0
\label{uni4}
\ff
so as to maintain the unimodular condition, (\ref{uni2}).  

Now we can write the gauge fixing condition in ``covariant" form as, 
\f
{\cal F}_b= \partial^a h_{ab} 
\ff

We can then write the path integral in terms of $h_{ab}$
\f
Z[J^{ab}] = e^{ W [J]^{ab}]}= \int dh_{ab} d\Psi  \delta (\mbox{ gauge fixing})  Det_{FP}e^{\imath( S^{uni}+ \int_{\cal M} h_{ab}J^{ab})}
\ff
where we have introduced the trace-free external current $J^{ab}$.  

We now follow the usual lagrange transform procedure to construct the effective action as a function of background fields $<h_{ab}>$.
\f
{\cal S}^{eff}[ <h_{ab}>]  = W[J] - \int <h_{ab}> J^{ab}
\ff
where 
\f
<h_{ab}>= \frac{\delta W}{\delta J^{ab}}|_{J=0}= \frac{1}{Z}  \int dh_{ab}  d\Psi  \delta (\mbox{ gauge fixing})  Det_{FP}  h_{ab} 
e^{\imath( S^{uni}+ \int_{\cal M} h_{ab}J^{ab})} |_{J=0}
\ff
Because of the unimodular condition (\ref{uni2}), we see directly that the background field is also trace free
\f
\delta^{ab} <h_{ab}> = 0 
\label{quni}
\ff
We can then write $< \bar{g}_{ab}>= exp <h_{ab}> $, with $det< \bar{g}_{ab}>=-1$ so that we write the effective action in terms of a unimodular
background field, $< \bar{g}_{ab}>$.  It is easy to see that in perturbation theory we will have, as usual, 
\f
{\cal S}^{eff}[ <\bar{g}_{ab}>] = S^{uni} ( <\bar{g}_{ab}>, \phi, \psi ) +\hbar \Delta S ( <\bar{g}_{ab}>, \phi, \psi )
\label{Seff}
\ff
We finally see the lesson, which is that the quantum effective action is also a function of a unimodular metric.  Therefor
the quantum equations of motion are functions of the unimodular metric.  

\section{Does unimodular gravity solve the second cosmological constant problem?}

We have seen that unimodular gravity may be quantized preserving the unimodularity condition.  It then offers a clean solution
to the first of the three cosmological constant problems.  But does it have anything to say about the second and third problems?   
This question has been discussed by Ng and van Dam \cite{nguni}
and Sorkin and collaborators\footnote{The latter propose an argument based on causal set theory for the magnitude of the
cosmological constant.  The argument appears to depend on specific details of causal sets, and there appears no relation to the
results here.}\cite{sorkinuni} and others.  

Let us review the proposal of Ng and van Dam \cite{nguni}
for canceling the cosmological constant.  They conjecture that the path integral for unimodular gravity can be written as 
\f
Z^{NvD} = \int d \mu (\Lambda ) Z^{GR}(\Lambda )
\label{NvD1}
\ff
where $Z^{GR}(\Lambda)$ is the partition function for general relativity for some value of the cosmological constant and
$d\mu (\Lambda )$ is some measure factor.  From this form they argue that in the semiclassical approximation for pure gravity, in which 
$S^{Einstein}(\Lambda ) \approx - \Lambda Vol/4\pi G$ where $Vol= \int_{\cal M} \sqrt{-g} $,  this is dominated by, formally,
\f
Z^{NvD} = \int d \mu (\Lambda )  dg_{ab} e^{-  \imath \frac{\Lambda}{4\pi G} Vol }
\label{NvD2}
\ff
they argue that the stationary phase approximation will be dominated by solutions where $\Lambda =0$.   

At first it is not clear that  this precise argument can be realized in the present context, because the ansatz (\ref{NvD1}) does not appear to be a form of the path integral for unimodular gravity.  For one thing, there is in the final gauge fixed partition function
(\ref{partition5b}) no integral over the cosmological constant.  
Nonetheless, it turns out we can reproduce the argument of Ng and van Dam in the present context.  

We go back to 
(\ref{partition1}) and write (being careful to keep in mind that the constraints are for each time and space point).
\f
\prod_{x^a} \delta ({\cal S}_i) = \int \prod_{x^a} d\lambda (x^i,t) \delta (h_0 + 4\pi G \rho  +  \lambda ) \delta (\partial_i \lambda )
\ff
We here have introduced a {\it spactime field} $\lambda (x^i, t).$  We can then introduce the lapse function
to write this as 
\f
\delta ({\cal S}_i) = \int \prod_{x^a} d\lambda (x^i,t) dN  \delta (\partial_i \lambda ) 
e^{\imath \int_{\cal M} N \sqrt{q}  (h_0+ 4\pi G \rho + \lambda ) }
\ff
We insert this into (\ref{partition1}) giving us
\begin{eqnarray}
Z &=&  \int  dq_{ij} dN d N^i d\lambda d \tilde{\pi}^{kl} d\tilde{a}^a dP_a  d\Psi dP_\Psi \delta (\mbox{ gauge fixing})  Det_{FP}
\nonumber \\
 & &\times  exp \  \imath \int_{\cal M}  \left (   \tilde{\pi}^{jk}\dot{q}_{jk}  +(\partial_a \tilde{a}^a)(h_0 +4 \pi G \rho )   + N^i {\cal D}_i
+N \sqrt{q}  (h_0+ 4\pi G \rho + \lambda )   \right ) 
 \label{partition11}
\end{eqnarray}
Recalling that the spacetime manifold $\cal M$ is spatially compact and we can ignore boundary terms in time we can
write this as 
\begin{eqnarray}
Z &=&  \int  dq_{ij} dN d N^i d\lambda d \tilde{\pi}^{kl} d\tilde{a}^a    d\Psi dP_\Psi   \delta (\mbox{ gauge fixing})  Det_{FP}
\nonumber \\
 & &\times  exp \  \imath \int_{\cal M}  \left (   \tilde{\pi}^{jk}\dot{q}_{jk}    + N^i {\cal D}_i
+(N \sqrt{q} +(\partial_a \tilde{a}^a))  (h_0+ 4\pi G \rho + \lambda )  +\tilde{a}^a \partial_a \lambda  \right ) 
 \label{partition12}
\end{eqnarray}
We can now shift the integration over $N$ by
\f
N\rightarrow N^\prime = N +\frac{\partial_a \tilde{a}^a}{\sqrt{q}}
\ff
We write $\lambda (x,t) = \Lambda (t) + \delta \lambda (t,x)$, with $\int_\Sigma \sqrt{q} \delta \lambda (t,x) =0$
and integrate over the spatial fluctuations $\delta \lambda (t,x)$ to find
\begin{eqnarray}
Z &=&  \int  \prod_t d\Lambda (t) \int \prod_{x^i} dq_{ij} dN d N^i d \tilde{\pi}^{kl} d\tilde{a}^a  d\Psi dP_\Psi  \delta (\mbox{ gauge fixing})  Det_{FP}
\nonumber \\
 & &\times  exp \  \imath \int_{\cal M}  \left (   \tilde{\pi}^{jk}\dot{q}_{jk}    + N^i {\cal D}_i
+N \sqrt{q}  (h_0+ 4\pi G \rho + \Lambda )  + \tilde{a}^0 \dot{ \Lambda}  \right ) 
 \label{partition13}
\end{eqnarray}
We now trivially integrate over the $\tilde{a}^i$.  The integral over $\tilde{a}^0$ yields a delta function 
$ \delta (\dot{\Lambda})$.  We again write $\Lambda (t) = \bar{\Lambda} + \delta \Lambda (t)$, where
again $\int dt \delta \Lambda (t)=0$.  The integral over the momenta $\tilde{\pi}^{ij}$ and $ dP_\Psi $can finally be done, yielding
\begin{eqnarray}
Z &=&  \int   d\bar{\Lambda} \int \prod_{x^a} dg_{ab}  d\Psi   \delta (\mbox{ gauge fixing})  Det_{FP}
\nonumber \\
 & &\times  exp \  \imath \int_{\cal M} \sqrt{-g} \left ( R + 2\bar{\Lambda} + {\cal L}^{matter}   \right ) 
 \label{partition14}
\end{eqnarray}
Thus we arrive at the Ng-Van Dam form (\ref{NvD1}).   From here we proceed as follows.  We evaluate this
only at the semiclassical approximation in which case it is 
\f
Z \approx   \int   d \Lambda \sum_{g_{ab}, \Psi}  
  exp \  \imath \int_{\cal M} \sqrt{-g} \left ( -\frac{\Lambda}{4\pi G} + ({\cal L}^{matter} -\frac{T}{2}) \right ) 
 \label{partition15}
\ff
where the sum is over pairs $(g_{ab}, \Psi )$ which solve of the Einstein equations with each value of $\Lambda$.  

In the stationary phase approximation this is dominated by histories in which 
\f
\frac{\Lambda }{4\pi G} \approx \frac{\int_{\cal M} \sqrt{-g} ( {\cal L}^{matter} -T )  }{Vol}= <( {\cal L}^{matter} -\frac{T}{2} ) >
\ff
For perfect fluids, ${\cal L}^{matter} = P,$ the pressure, so neglecting that we find that the most likely value of the 
cosmological constant is 
\f
\frac{\Lambda }{2\pi G} \approx \frac{\int_{\cal M} \sqrt{-g} \rho }{Vol}
\label{relation}
\ff

Thus we arrive at a relation between the cosmological constant and the average, over the spacetime volume over the
history of the universe, of the energy density.

It is diffiucult to know how to evaluate (\ref{relation}) because it is so formal, and because to make real sense of it requires
a precise measurement theory for quantum cosmology, which we lack.  Indeed, it is not clear we could give a precise
operational meaning to the average of a quantity over the whole spacetime volume of the universe.  

But suppose we make the simplest assumption, which is to employ the assumption of mediocrity\footnote{Note that we
are employing the notion of typicality or mediocrity within a single universe which, however problematic, is less problematic
than employing it in a multiverse within which our universe is far from typical.}.
We then posit that our present situation is typical, so that we can replace the average value by the present
value.  This gives us 

\f
\frac{\Lambda }{ G} \approx 2 \pi \rho
\ff
which is remarkably close for such a speculative argument.

Before closing, we can note that another route to the same result is as follws.  Let us start with
(\ref{partition5b}) and write in the semiclassical approximation
\begin{eqnarray}
\langle \Lambda \rangle  &=& \frac{1}{Z}  \int  dg_{ab}  d\psi \delta (\sqrt{-g} -\epsilon_0 )
\delta (\mbox{ gauge fixing})  Det_{FP} \sqrt{-g} 
\nonumber \\
 & &\times \Lambda (g_{ab}, \psi)     exp \  \imath \int dt \int_\Sigma \left ( \epsilon_0 ( \frac{1}{8\pi G}R +{\cal L}^{matter} ) \right ) 
 \label{partition5c}
\end{eqnarray}
where we denote by $ \Lambda (g_{ab}, \psi) $ the cosmological constant as determined for solutions by (\ref{Lambda1})
\f
 \Lambda (g_{ab}, \psi) = -\frac{1}{4} (R +4 \pi G T) 
\label{Lambda2}
\ff
In the semiclassical approximation we have, using again (\ref{Lambda1}) and the unimodular condition ({\ref{uni2})
\begin{eqnarray}
\langle \Lambda \rangle  &\approx & \frac{1}{Z}  \int  dg_{ab}  d\psi \delta (\sqrt{-g} -\epsilon_0 )
\delta (\mbox{ gauge fixing})  Det_{FP} \sqrt{-g} 
\nonumber \\
 & &\times \Lambda (g_{ab}, \psi)     exp \  \imath  \left ( -\frac{\Lambda  (g_{ab}, \psi)}{2\pi G} Vol  
 +\int_{\cal M} \sqrt{-g} ( {\cal L}^{matter} -\frac{T}{2}  ) \right ) 
 \label{partition5c}
\end{eqnarray}

From here the argument is the same as above.  

\section{Conclusions}

The results here indicate that unimodular gravity does in fact provide a solution to the first of the cosmological constant problems.  Furthermore, we
showed that a solution to the second cosmological constant problem, first advocated by Ng and van Dam, does indeed arise
in a semiclassical approximation to the path integral that defines the quantum theory.  We even got a bonus, which is a possible
hint about the third, coincidence problem.  

Here is one perspective these results suggest. Since the different forms of unimodular gravity have the same field equations as general relativity, the quantization of any of them can be considered a quantization of general relativity.    But 
one can argue that any consistent quantum theory of gravity must solve the first cosmological constant problem as  a matter of self-consistency.  For only if that problem is solved do the solutions
of the theory leave the quantum domain, allowing a classical limit to exist where the predictions of general relativity might be restored.  
Given these points, one  can conclude that, if the quantizations of unimodular gravity are the only quantizations of general relativity that solve the first cosmological constant problem, they are also the only consistent quantizations.  

We close with a few comments on these results

\begin{itemize}

\item{}We must emphasize that the results here are formal, as they assume 
the existence of the partition function defined by functional integrals.  Thus,  these results must be
confirmed within a framework in which the path integral has been shown to exist.  Two approaches which resolve the ultraviolet
problems of quantum gravity at a non-perturbative and background independent level are causal dynamical triangulations and
spin foam models.  Thus, the next step must be to re-express these results in those frameworks.  
  A first step in this direction is the casting of unimodular gravity in Ashtekar variables \cite{bombelliuni}.
  
  \item{}It is also  worthwhile to investigate whether these results can be realized in a string theoretic or supersymmetric
  context. Here one can begin with unimodular forms of supergravity, shown to exist in \cite{superuni}.  

\item{} One may ask whether, when regularization is fully taken into account, there might arise anomalies in the condition that the partition function depends on the unimodular metric.    We can make two comments about this.  First, the condition that the gauge fixed path integral (\ref{partition5b}) involves an integral only on metrics with fixed determinants is not analogous to the situation of the trace anomaly.  First of all,  the results here do not depend on the action having an overall scale invariance, globally or locally.  There may be mass terms, and there may also be a conformal anomaly, these do not effect the conclusions.  One may 
nonetheless ask if an anomaly analogous to the trace anomaly for the energy momentum tensor may spoil these results.  That is, could there be anomalous contributions to $Tr< h_{ab} > $ in (\ref{quni}), the same way there are in quantum field theories anomalous contributions to  $Tr <0| T_{ab} | 0 >$?  The point of section 4 is that they do not appear to arise.  The anomaly in the trace of the expectation value of the energy momentum tensor arises from the need to break scale invariance to regulate an operator product in the definition of the energy-momentum tensor.  As $h_{ab}$ is not an operator product no such issue arises here.  

\item{}Furthermore, the condition of unimodularity can be imposed within a regulated form of the partition function for general relativity. 
It is imposed directly in the causal set approach to quantum gravity, as emphasized by Sorkin\cite{sorkinuni}, which is an approach where there is one quantum event per spacetime Planck volume.  It is also imposed in the dynamical triangulations approach, because the regulated path integral is constructed by integrating over piecewise flat manifolds, each constructed by gluing simplices together of fixed geometry and hence fixed volume.  And it can be easily imposed in  spin foam models, as will be discussed elsewhere.  So the path integral (\ref{partition5b}) can be regulated in these different ways while maintaining the unimodularity condition.  This is dissimilar to the 
cases where anomalies arise in conditions such as conformal invariance and chiral gauge invariance, where the invariances are broken by the regulators needed to define the path integral. Thus, there is no reason to expect the unimodularity condition will not survive translation of (\ref{partition5b}) into a fully regulated nonperturbative definition of the path integral. 

\item{} There are claims that the weight of quantum fluctuations has been measured already in the contribution to atoms coming 
from the Lamb shift.  This does not contradict the present results as those contributions are not of the form of $\bar{g}_{ab} V$.  

\item{} It has been said that there is a strong case for multiverse cosmologies and the anthropic principle, because they provide the
only known solution to the cosmological constant problem.  To the extent that proposals such as Afshordi's\cite{niayesh} and the
present solve one and perhaps all of the cosmological constant problems, this argument cannot be made.

\item{}Sorkin\cite{sorkinuni} and Unruh \cite{unruhuni,unruhwald}  have pointed out that unimodular gravity has a non-vanishing hamiltonian and hence evolves
quantum states in terms of a global time, given by an analogue of the Schrodinger equation. 
Within the present framework this Schrodinger equation arises from the quantization of the global constraint equation (\ref{ham}).  
This, as Sorkin, Unruh and others have argued, offers a new perspective
on the problem of time in quantum cosmology.     There are also other, independent, reasons to be interested in the notion of a global time in quantum cosmology\cite{withroberto}.  Given the results discussed here, these suggestions deserve greater consideration.

\end{itemize}

\section*{ACKNOWLEDGEMENTS}

I am grateful first of all to Niayesh Afshordi for several discussions of his model which inspired the present one as well as pointing out a mistake in an early draft.    Thanks are also due to Giovanni Amelino-Camelia, Laurent Freidel, Sabine Hossenfelder and Carlo Rovelli for crucial comments on a draft, as well as to Rafael Sorkin for conversations and Marc Henneaux for correspondence.  Research at Perimeter Institute for Theoretical Physics is supported in part by the Government of Canada through NSERC and by the Province of
Ontario through MRI.

\appendix.

\section{More formulations of unimodular gravity}

For the interested reader I describe two more forms of unimodular gravity.

\subsection{The inverse Henneaux Teitelboim action}

We can reverse the trick of Henneaux and Teitelboim.  Instead of (\ref{HT}) we consider
\f
S^{HT^{-1} } (\bar{g}_{ab}, a_{abc},\phi,...)  = \int_{\cal M} \tilde{b} \left ( -\frac{1}{8\pi G}  (  \bar{g}^{ab} R_{ab} + \phi)  + {\cal L}^{matter} \right )
-\frac{1}{8\pi G}  \phi \sqrt{-g}
\label{IHT}
\ff
The equations of motion include again the unimodular condition (\ref{uni5}).  But instead of the Einstein equations we have
\f
\tilde{b} R_{ab}+ \frac{1}{2}\sqrt{-g}g_{ab} \phi = \tilde{b} 4\pi G E_{ab}
\label{IHT1}
\ff
Together with the condition coming from the variation of $\tilde{a}^a$, 
\f
\partial_a \left ( R+ 8 \pi G {\cal L}^{matter} +\phi \right ) =0
\ff
We can now define
\f
R+ 8 \pi G  {\cal L}^{matter} +\phi =-2 \Lambda
\ff
a constant, from which  we again return to the Einstein 
equation (\ref{HT2}).   Again we have the symmetry (\ref{modify}) together with $\Lambda \rightarrow \Lambda -16 \pi G C$.

\subsection{Another  reformulation of unimodular gravity}

Just to show the variety of formulations that are possible, I give here yet another formulation of unimodular gravity.  This differs
from the others in that the equations of motion are not precisely Einstein's equations.   

In this formulation we  drop the $\phi$ field altogether but we  get the unimodular condition (\ref{uni5}) in  a subset of solutions to equations of motion.  
We can take the solution of the equation (\ref{uni2}) and plus it back
into the action (\ref{HT}), replacing $\sqrt{-g}$ by $\tilde{b}$ to find
\f
S^{4} (\bar{g}_{ab}, a_{abc},..._)  = \int_{\cal M} \tilde{b} \left ( -\frac{1}{8\pi G_0}   \bar{g}^{ab} R_{ab} + {\cal L}^{matter} \right )
\label{uniaction2}
\ff
where by $\bar{g}_{ab}$ is meant the metric restricted by the condition (\ref{uni2}).  We work out the equations of motion 
\f
(\partial_a \tilde{a}^a ) \left ( R_{ab}-\frac{1}{4}\bar{g}_{ab} R \right ) + 2 (W_{ab}  - \frac{1}{4}\bar{g}_{ab} W )= (\partial_a \tilde{a}^a ) 
4 \pi G_0 \left ( E_{ab}-\frac{1}{4}\bar{g}_{ab} E \right )
\label{EOM1}
\ff
where $W_{ab}$ is the tensor
\f
W_{ab}= \nabla_a\nabla_b \tilde{b} 
\ff
which is nonvanishing because for generic solutions $\tilde{b}$ and $\sqrt{-g}$ will not be proportional.  

Variation of the action (\ref{uniaction2}) by $a_{abc}$ gives rise to a second set of equations of motion.
\f
\partial_a \left (  R+ 8 \pi G {\cal L}^{matter}   \right ) =0
\label{EOM2}
\ff
We can then define for each solution 
\f
R+ 8 \pi G  {\cal L}^{matter} = -4\Lambda
\label{lambda}
\ff
a spacetime constant.  This replaces the trace of the Einstein equations so that now, 
\f
G_{ab}- \bar{g}_{ab} \Lambda + \frac{2}{\tilde{b}} (W_{ab}- \frac{1}{4}\bar{g}_{ab} W )  =   
4 \pi G_0  T_{ab} - \pi G_0  g_{ab} \left ( T- 4{\cal L}^{matter}    \right ) 
\label{EOM3}
\ff

Since (\ref{EOM1}) is trace-free there is a missing equation. It is replaced by 
another equation which arises from taking
the covariant divergence of the Einstein equations gives 
\f
\nabla_b \left ( \frac{1}{\tilde{b}} W_{a}^b \right ) - \frac{1}{4} \nabla_a\left ( \frac{1}{\tilde{b}} W \right )  =
-\pi G \nabla_a  \left ( T- 4{\cal L}^{matter}    \right ) 
\label{EOM4}
\ff
A solution to this for vacuum is of course
\f
\nabla_a \tilde{b} =0
\ff
which implies that 
\f
\tilde{b}= \epsilon_0 \cdot constant = \sqrt{-g}\cdot constant
\label{recovery}
\ff
but there will be other solutions.

Thus, we see that for the vacuum case, and only for the vacuum case,  solutions to Einstein's equations are recovered.  $\Lambda$ is, however, a constant of motion, so that
the solutions to this theory include solutions to Einstein's equations for any value of $\Lambda$.   In the case that there is matter, this theory is different from general relativity, as is the theory of Afshordi\cite{niayesh}.  Work will need to be done to see if it  agrees with the known tests of general relativity, as the Afshordi theory appears to do.  

\subsection{Other forms of the partition function}

At the semiclassical level we can make the substitution of the argument of the delta function in the action to find
\begin{eqnarray}
Z &\approx &  \int  dg_{ab}  d\tilde{a}^a \delta (\sqrt{-g} -\tilde{b} )
\delta (\mbox{ gauge fixing})  Det_{FP}  \sqrt{-g} 
\nonumber \\
 & &\times  exp \  \imath \int dt \int_\Sigma \left ( \sqrt{-g}  ( \frac{1}{8\pi G}R +{\cal L}^{matter} ) \right ) 
 \label{partition6}
\end{eqnarray}
Or we can proceed from (\ref{partition5}) to introduce the lagrange multiplier field $\phi$ to exponentiate the unimodular
constraint to find
\begin{eqnarray}
Z &=&  \int  dg_{ab}  d\tilde{a}^a d\phi
\delta (\mbox{ gauge fixing})  Det_{FP}  \sqrt{-g} 
\nonumber \\
 & &\times  exp \  \imath \int dt \int_\Sigma \left ( \tilde{b} ( \frac{1}{8\pi G}R +{\cal L}^{matter} +\phi ) -\phi \sqrt{-g} \right ) 
 \label{partition7}
\end{eqnarray}
giving us for an action the inverse-Henneaux-Teitelboim version of the theory (\ref{IHT}).  Or, we can obtain the 
Henneaux-Teitelboim from from (\ref{partition6}) by exponentiation, giving us
\begin{eqnarray}
Z &\approx &  \int  dg_{ab}  d\tilde{a}^a d\phi
\delta (\mbox{ gauge fixing})  Det_{FP}  \sqrt{-g} 
\nonumber \\
 & &\times  exp \  \imath \int dt \int_\Sigma \left ( \sqrt{-g} ( \frac{1}{8\pi G}R +{\cal L}^{matter} +\phi ) -\phi  \tilde{b} \right ) 
 \label{partition7}
\end{eqnarray}

\end{document}